\documentclass[pre,twocolumn,nofootinbib,superscriptaddress]{revtex4}

\usepackage{amsmath,amssymb}
\DeclareMathOperator*{\argmax}{arg\!max}
\usepackage{graphicx}
\usepackage{bm}
\usepackage{algorithm}
\usepackage[noend]{algpseudocode}
\usepackage{color}
\usepackage[colorlinks,linkcolor=blue,anchorcolor=blue,citecolor=blue]{hyperref}
\usepackage[T1]{fontenc}
\definecolor{blue}{rgb}{0.0,0.0,0.0}

\begin{document}

\title{Change-point detection in anomalous-diffusion trajectories utilising
machine-learning-based uncertainty estimates}

\author{Henrik Seckler}
\affiliation{Institute for Physics \& Astronomy, University of Potsdam, 14476 Potsdam-Golm, Germany}
\author{Ralf Metzler}\email{rmetzler@uni-potsdam.de}
\affiliation{Institute for Physics \& Astronomy, University of Potsdam, 14476 Potsdam-Golm, Germany}
\affiliation{Asia Pacific Centre for Theoretical Physics, Pohang 37673,
Republic of Korea}

\date{\today}

\begin{abstract}
When recording the movement of individual animals, cells or molecules one will
often observe changes in their diffusive behaviour at certain points in time
along their trajectory. In order to capture the different diffusive modes
assembled in such heterogeneous trajectories it becomes necessary to segment
them by determining these change-points. Such a change-point detection can
be challenging for conventional statistical methods, especially when the
changes are subtle. We here apply \emph{Bayesian Deep Learning} to obtain
point-wise estimates of not only the anomalous diffusion exponent but also the
uncertainties in these predictions from a single anomalous diffusion trajectory
generated according to four theoretical models of anomalous diffusion.
We show that we are able to achieve an accuracy similar to single-mode
(without change-points) predictions as well as a well calibrated uncertainty
predictions of this accuracy. Additionally, we find that the predicted
uncertainties feature interesting behaviour at the change-points leading us
to examine the capabilities of these predictions for change-point detection.
While the series of predicted uncertainties on their own are not sufficient
to improve change-point detection, they do lead to a performance boost when
applied in combination with the predicted anomalous diffusion exponents.
\end{abstract}

\maketitle

\section{Introduction}

\begin{figure*}
\centering
\includegraphics[width=\linewidth]{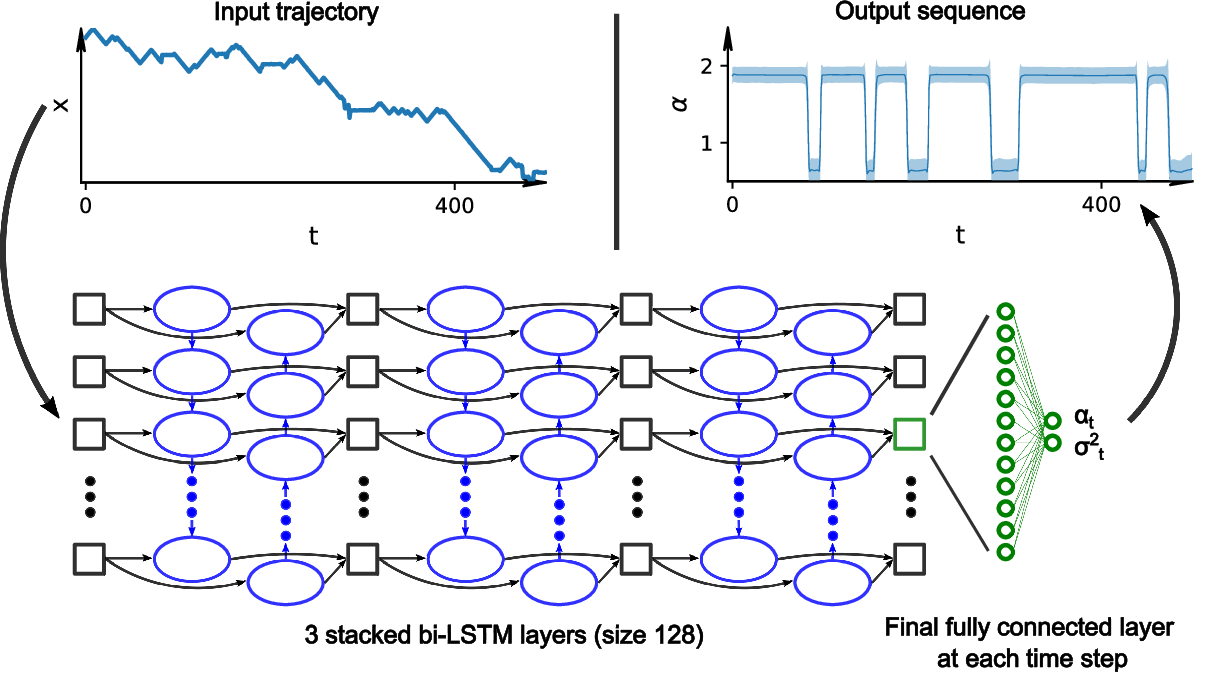}
\caption{Visualisation of the machine learning model. The model provides a 
point-wise prediction of the anomalous diffusion exponent and its uncertainty 
given a single-particle diffusion trajectory. The output sequence in the upper 
right panel indicates the predicted error as a shaded area of $\pm \sigma$ 
around the mean. After the trajectory is converted to increments and 
normalised, it is fed into a stacked \textit{bidirectional 
long-short-term-memory} (bi-LSTM)~\cite{hochreiter1997long} neural network 
consisting of 3 layers of size 128. This network returns a time series of 
vectors of size 128, 
each of which is again passed through a single-layer fully connected network 
to produce the time series of estimates for the mean of the anomalous 
diffusion exponent $\alpha_t$ and its variance $\sigma^2_t$. The final 
predicted output time-series is obtained by combining the output of multiple such 
networks,  sampled from a \emph{Multi-SWAG} 
estimate~\cite{wilson2020bayesian}.}
\label{fig_architect}
\end{figure*}

In ancient Rome, around 60 BC, the philosopher Lucretius gives a remarkable
description in his poem \textit{"De rerum natura"} (\textit{"On the Nature
of Things"}), observing the motion of dust particles in a ray of light in
a stairwell~\cite{lucretius60rerum}. While what he called the "dancing" of
particles is predominantly caused by air currents, he provided a surprisingly
accurate explanation for the much later understood phenomenon of Brownian
motion~\cite{brown1828xxvii,einstein1905molekularkinetischen}. He postulated
that the motion must be caused by smaller unseen particles colliding into
larger ones until cascading to the visible motion.

Mathematically such a motion is modelled as a path of successive random
steps, called a random walk~\cite{pearson1905problem,weissbook}. As
long as these increments $\Delta x(t)$ are independently, identically
distributed with finite variance, they will, under the central limit theorem
(CLT), lead to normal diffusion~\cite{mises1919fundamentalsatze}. The
prime example of this is the aforementioned Brownian motion, as
observed by Robert Brown in 1827, when tracking pollen granules suspended in
water~\cite{brown1828xxvii,einstein1905molekularkinetischen,smoluchowski1906kinetischen,sutherland1905lxxv,langevin1908theorie}.
Amongst others, normal diffusion entails a Gaussian probability density
function (PDF) of the position $x(t)$ with a linear growth of the mean
squared displacement (MSD)~\cite{Kampen1981stochastic,hughes1995random}\footnote{We
here consider the one-dimensional case. Future work will study the performance of
the developed algorithm in higher dimensions.}
\begin{equation}
\langle x^2(t)\rangle\sim2Dt.
\end{equation}

In reality however, many experiments show a non-linear growth of the 
MSD~\cite{bouchaud1990anomalous,metzler2000random,golding2006physical, manzo2015weak, krapf2019spectral,stadler2017non,kindermann2017nonergodic,sokolov2012models,hofling2013anomalous,horton2010development,tolic2004anomalous,jeon2013anomalous,saxton1994anomalous,saxton2001anomalous,pt,burov2011single,ernst2014probing,grossmann2024non}, 
indicating that one or multiple conditions of the CLT are broken. Often these 
systems feature a power-law growth 
\begin{equation}
\langle x^2(t)\rangle\sim2K_\alpha t^\alpha
\end{equation}
of the MSD,
which is referred to as anomalous diffusion with anomalous diffusion exponent 
$\alpha$ and generalised diffusion constant $K_\alpha$ \cite{pt,metzler2000random,
bouchaud1990anomalous}. A motion with 
slower-than-normal growth of the MSD (with $0<\alpha<1$) is called 
subdiffusive, whereas for $\alpha>1$ it is referred to as superdiffusive. 
Mathematically, anomalous diffusion may be achieved in many ways, all breaking 
at least one condition of the 
CLT~\cite{bouchaud1990anomalous,metzler2000random,metzler2014anomalous}. Some 
examples of this include long-range 
correlated increments in fractional Brownian motion 
(FBM)~\cite{mandelbrot1968fractional}, non-identically distributed increments 
in scaled Brownian motion (SBM)~\cite{jeon2014scaled,lim2002self} or 
increment-distributions with infinite variance in L{\'e}vy
flights~\cite{chechkin2008introduction,levy1937theorie}. Additionally, 
introducing waiting times between jumps in a so called continuous time random 
walk (CTRW) allows for further options to reach anomalous 
diffusion~\cite{montroll1965random,hughes1981random,weissman1989transport,shlesinger1986levy,zaburdaev2015levy}. 
A more detailed description of these models is 
provided in subsection \ref{subsec_anodif}.

As each of these models corresponds to a different cause of anomalous 
diffusion, determining the model best fitting experimental data can yield 
insights into the underlying diffusive mechanisms. Furthermore one may wish to 
determine the associated parameters such as the anomalous diffusion exponent 
$\alpha$ or the generalised diffusion constant $K_\alpha$. The experimental data provided 
for these tasks usually consist of single particle trajectories, tracking the 
movement of molecules in 
cells~\cite{elf2019single,cherstvy2019non,hofling2013anomalous,horton2010development, tolic2004anomalous,leijnse2012diffusion,codling2008random}, 
animals~\cite{okubo1986dynamical, vilk2022unravelling,bartumeus2005animal} or 
even stock prices~\cite{malkiel1999random, plerou2000economic}.

The tools to decipher the stochastic processes underlying measured data and the
associate parameters are various, 
ranging from the use of several 
statistical observables~\cite{metzler2009analysis,magdziarz2009fractional,burov2011single,metzler2019brownian,vilk2022classification,condamin2007first,slezak2019codifference,aghion2021moses,meyer2022decomposing,sposini2020universal,sposini2022towards,burnecki2015estimating,wylomanska2015codifference} 
to Bayesian inference~\cite{thapa2018bayesian,thapa2022bayesian,park2021bayesian} 
or machine learning~\cite{seckler2022bayesian,seckler2023machine,gentili2021characterization,granik2019single,kowalek2019classification,firbas2023characterization,al2022classification,bo2019measurement,gajowczyk2021detection,argun2021classification,garibo2021efficient,li2021wavenet,verdier2021learning,seckler2024multifractal,munoz2020single}. 
The latter demonstrated its potential in the \emph{Anomalous Diffusion (AnDi) 
Challenge} held in 2020~\cite{munoz2020anomalous,munoz2021objective}. In this 
challenge participants were tasked with the determination of the anomalous 
diffusion model and its anomalous diffusion exponent $\alpha$ from 
computer-generated single-particle trajectories. Especially when data is 
sparse, machine-learning models tended to outperform the conventional ones.

In a recent paper we showed that using \emph{Bayesian Deep Learning}, one can 
expand the machine learning solutions to the \emph{AnDi-Challenge} to include 
an uncertainty estimate~\cite{seckler2022bayesian,seckler2023machine}. This 
estimate allows for an easier assessment whether predictions are reliable and 
improves the interpretability of the algorithm. 

We here aim to explore the applicability of this machine-learning model to 
diffusion trajectories in which the walker may change between diffusive modes. 
Such a behaviour may occur, for instance, when the tracked animal or cell 
switches between a searching and feeding state, as an effective behaviour in an
heterogeneous environment or due to transient interactions with other 
objects~\cite{vilk2022unravelling,yin2018detection,saha2015diffusion,bag2015plasma,low2011erbb1, moldenhawer2022spontaneous, monasson2014crosstalk}. 
If one desires to differentiate between these diffusive modes 
it is imperative to be able to segment the trajectory by identifying the 
points at which the changes 
occur~\cite{wagner2017classification,persson2013extracting,stanislavsky2024fractional,janczura2021identifying,arcizet2008temporal}. 

Applications of machine-learning techniques to this problem are explored in 
the second \emph{AnDi-Challenge}~\cite{munoz2023quantitative}, whose results 
are expected to be published in the near future; note however that the setup 
differs strongly from the one described herein and can thereby not be compared 
directly. Already published works suggest that a point-wise prediction, in 
which one provides a prediction of the diffusion parameters at each point in 
time, using machine learning may be a promising 
solution~\cite{requena2023inferring,qu2024semantic}. 

In this work we expand on previous works by training a \emph{Bayesian Deep 
Learning} model to provide a point-wise prediction of not only the anomalous 
diffusion exponent but also its uncertainty in this prediction for a given 
single-particle trajectory, see Fig.~\ref{fig_architect} for a schematic 
representation. We subsequently aim to extract the change-points between the 
diffusive modes from the sequences of predicted anomalous diffusion exponents 
and their uncertainties, that exhibit a peak around the change-point.
\textcolor{blue}{Apart from the obvious advantage from using an additional 
time series, we also expect this to improve change-point detection in cases
when the anomalous diffusion exponent itself does not change significantly,
which could lead to a failed change-point detection if only the point-predictions
for $\alpha$ were used. However, even for such a small change in $\alpha$ the
uncertainty will typically show a more pronounced signal and thus improve
detection.}

The paper is structured as follows. We begin with an explanation of the 
utilised anomalous diffusion data set (subsection \ref{subsec_anodif}), machine 
learning procedures (subsection \ref{subsec_machproc}) as well as evaluation 
metrics (subsection \ref{subsec_avametr}). Subsequently we present the obtained
results in section \ref{sec_results}---this includes the 
performance of the point-wise prediction (subsection \ref{subsec_wholeseq}) as 
well as a detailed evaluation of change-point extraction methods (subsection 
\ref{subsec_cpextract}), followed by a short consideration of parameter 
predictions for the diffusive modes (subsection \ref{subsec_twomodepred}). We 
finish with a discussion of the results in section \ref{sec_discuss}.

\section{Methods}
\label{sec_methods}
\subsection{Anomalous diffusion data set}
\label{subsec_anodif}
The data set is generated utilising 4 different diffusion models, all yielding 
anomalous diffusion with ${\langle x^2(t) \rangle \sim 2 K_\alpha t^\alpha}$. 
Namely, we use fractional Brownian motion (FBM), scaled Brownian motion (SBM), 
continuous time random walk (CTRW) and L{\'e}vy Walk (LW).

FBM is the paradigmatic model for breaking the independence condition of the 
CLT. It is characterised by the covariance function
\begin{equation}
    \langle x(t)x(t+\tau)\rangle = K_\alpha(|t|^\alpha+|t+\tau|^\alpha-|\tau|^\alpha), 
\end{equation}
which constitutes a long-time correlation of the increments 
in the power-law form
\begin{equation}
\langle\Delta x(t)\Delta x(t+\tau)\rangle\sim\alpha(\alpha-1) K_\alpha \tau^{\alpha-2}
\end{equation}
for sufficiently large $\tau$, where $\alpha$ is the anomalous diffusion 
exponent and $K_\alpha$ is the generalised diffusion 
constant~\cite{mandelbrot1968fractional}. This correlation entails a 
persistent or anti-persistent motion for $\alpha>1$ (superdiffusion) or 
$\alpha<1$ (subdiffusion) respectively. 

SBM breaks the identity condition of the CLT. This is achieved by introducing 
a time dependent diffusivity $K(t)=\alpha K_\alpha t^{\alpha-1}$ in the 
Langevin equation
\begin{equation}
\frac{dx(t)}{dt}=\sqrt{2K(t)}\xi(t),
\end{equation}
where $\xi(t)$ is white, zero-mean Gaussian noise \cite{jeon2014scaled}.

In a CTRW both the jumps and the waiting times between those jumps are 
stochastic 
variables~\cite{montroll1965random,hughes1981random,weissman1989transport}. 
We here consider a CTRW with a heavy-tailed waiting time PDF $\psi(\tau)$, such
that $\psi(\tau)\propto\tau^{-1-\alpha}$ for 
large $\tau$ with scaling exponent $0<\alpha<1$. The resulting diverging mean 
waiting time $\int_0^\infty \tau\psi(\tau) d\tau = \infty$ leads to a 
subdiffusive behaviour. The spatial displacements are drawn from a zero-mean 
Gaussian PDF.

LWs can be considered as a special case of a
CTRW~\cite{shlesinger1986levy,zaburdaev2015levy}. As before it features a 
power-law waiting time PDF $\psi(\tau)\propto\tau^{-1-\kappa}$, but, instead 
of only jumping after a waiting time, the walker travels with a constant speed 
$v$ in one direction for the duration of one waiting time. After this time a 
new random direction is chosen. As shown in Ref.~\cite{zaburdaev2015levy} this 
leads to superdiffusion with anomalous diffusion exponent 
\begin{equation}
\alpha=\left\{\begin{array}{ll}2&\mbox{ if } 0<\kappa<1 \mbox{ (ballistic
diffusion)}\\3-\kappa & \mbox{ if } 1<\kappa<2 \mbox{ (superdiffusion)}.
\end{array}\right. 
\end{equation}
We use the \texttt{andi-data sets} Python package for the implementation of 
these models~\cite{munoz2021andichallenge}.

We generate multiple data sets consisting of different numbers $N$ of 
trajectories. Each trajectory consists of $T=500$ data points and will 
randomly switch between two diffusive modes, each following one of the four 
diffusion models described above. For each mode, we choose the anomalous 
diffusion exponent $\alpha$ from a uniform random distribution 
$\alpha\in\{0.05,0.1,\dots,2.0\}$, with restrictions based on the underlying 
diffusion model. Note that LWs may only be superdiffusive $\alpha\geq1$, 
CTRWs only subdiffusive $\alpha\leq1$, and ballistic ($\alpha=2$) FBMs are 
not considered here. Thus, in each simulated trajectory we simulate sequences
of two individual anomalous diffusion models, each with a given $\alpha$ and
$K_{\alpha}$. Each of the two diffusive modes is 
generated with an increment standard deviation of 1 and subsequently 
multiplied by a factor taken from a Gaussian distribution with mean 0 and 
variance 1, the square value of which is thereby equivalent to twice
the (short-time) diffusion constant.
As trajectories will be normalised later on, only the ratio of the diffusion 
constants of the two modes is of importance here. Note that the same diffusion 
model and/or parameters may be chosen for both modes. 

To generate the change-points between the modes we draw dwell times from an 
exponential distribution 
\begin{equation}
    \Psi(t) = \frac{1}{\beta}\times\exp\left(-\frac{t}{\beta}\right).
    \label{eq_dwell}
\end{equation}
The scale parameter $\beta$ is itself a random variable drawn for each trajectory 
from a reciprocal distribution $p(\beta)\propto 1/\beta$, but limited within 
$50<\beta<250$. This choice corresponds to the non-informative prior 
(Jeffrey's prior) for the exponential 
distribution~\cite{jeffreys1946invariant,jeffreys1998theory}, the limits were 
chosen to generate trajectories that feature a few but not too many 
change-points within the total trajectory length $T=500$.

All data are corrupted by white Gaussian noise with a signal to noise strength 
ratio randomly chosen from $\text{snr}\in\{1,2,10\}$. Given the trajectory 
$x(t)$, we obtain the noisy trajectory $\tilde{x}(t)=x(t)+\xi(t)$ with the 
superimposed noise
\begin{equation}
\xi(t) \sim \frac{\sigma_{\Delta x}}{\text{snr}} \mathcal{N}(0,1)\text{,}\label{eq_noise}
\end{equation}
where $\sigma_{\Delta x}$ is the standard deviation of the increment process 
$\Delta x (t) = x(t+1)-x(t)$. Note that if the diffusion constants of the two 
modes differ strongly, one will be much more affected by noise than the other, 
as would also be the case in real tracking experiments.

We generate a total of 4 data sets. A training data set consisting of 
$N=5\times10^6$ trajectories used to train the neural network, a validation 
data set of $N=1\times10^5$ for fine-tuning of the machine learning 
hyper-parameters, as well as two test data sets consisting of $N=5\times10^5$ 
trajectories each. The first is utilised to fine-tune the methods for 
change-point extraction, which are then applied to the second test data set to 
generate the results herein. 

To allow for general applicability independent of scale as well as in order to 
stabilise the training process, trajectories undergo a minimal amount of 
preprocessing. Concretely, each trajectory is converted into its increment 
series and normalised to a uniform standard deviation before being fed into 
the neural network.

\subsection{Machine-learning procedure}
\label{subsec_machproc}

\begin{figure*}
    \centering
    \includegraphics[width=\linewidth]{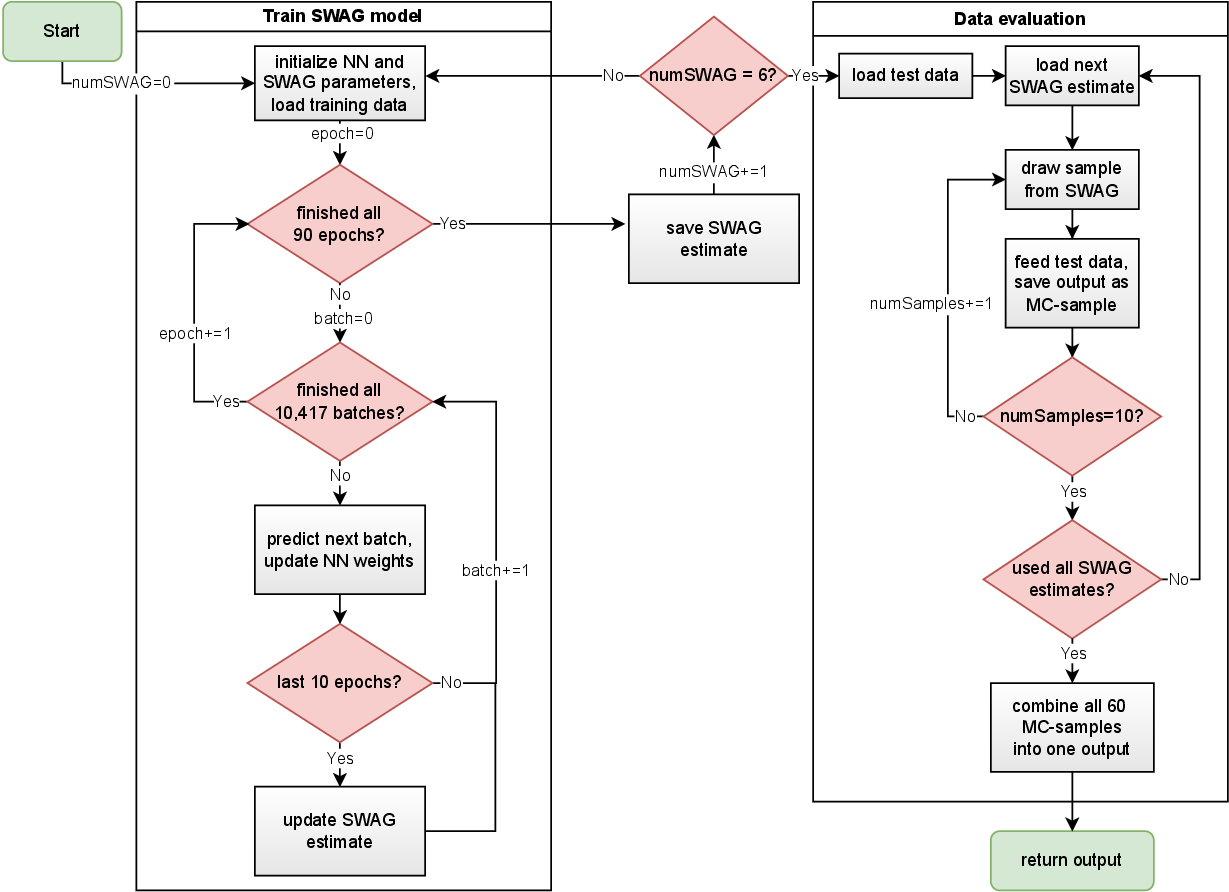}
    \caption{
    Flowchart of the \emph{multi-SWAG} procedure used for training and evaluation. The \emph{multi-SWAG}~\cite{wilson2020bayesian} model consist of 6 \emph{SWAG} models, which estimate a Gaussian posterior on the neural network (NN) weights during the last 10 epochs of the training loop. During data evaluation we sample 10 networks from each \emph{SWAG} estimate for a total of 60 Monte-Carlo-samples (MC-samples). These samples are combined into a single output prediction of means and uncertainties. Note that the final predicted uncertainties are the sum of the mean predicted variances and the variances of predicted means~\cite{kendall2017uncertainties,seckler2022bayesian}.}
    \label{fig_flowchart}
\end{figure*}

In recent years many neural network architectures for the analysis of 
anomalous diffusion trajectories have been 
proposed~\cite{munoz2021objective,seckler2023machine}. These range from fully 
connected~\cite{gentili2021characterization} to 
convolutional~\cite{granik2019single,kowalek2019classification,firbas2023characterization,al2022classification}, 
recurrent~\cite{bo2019measurement,gajowczyk2021detection,argun2021classification,garibo2021efficient,li2021wavenet} 
or even graph neural networks~\cite{verdier2021learning}. After some initial 
probing we here decided on a recurrent neural network as depicted in figure 
\ref{fig_architect}. The network consists of three stacked 
\textit{bidirectional long-short-term-memory} 
(bi-LSTM)~\cite{hochreiter1997long} layers of size 128, which return a time 
series of 128-dimensional vectors. At each time step $t$, these are passed 
through a fully connected layer, outputting estimates for mean $\alpha_t$ and 
variance $\sigma^2_t$ of the anomalous diffusion exponent. 

\textcolor{blue}{We choose the recurrent LSTM-architecture due to its
design for time-series prediction as well as its immediately accessible
output of a time-series of the same length as the input. Additionally,
this architecture has been successfully applied to anomalous diffusion
trajectories before \cite{argun2021classification,seckler2022bayesian}. We
added bidirectionality as the point-wise predictions pursued here should
heavily depend on future data points, which is not the case in a unidirectional
LSTM. The three layers of size 128 were chosen via trial-and-error, testing
larger networks until no further improvement was perceived.}
Note that we also 
tested a multi-headed attention architecture~\cite{vaswani2017attention} but 
found that this did not improve results noticeably for our setup, especially 
considering its significantly increased computational cost.

To find the optimal weights of the neural network we minimise a loss function 
on the training data set.
As the outputs include uncertainty estimates, the network cannot be trained by 
using an---otherwise standard for regression tasks---squared error loss. 
Instead we minimise the \textit{Gaussian negative log-likelihood loss}
\begin{equation}
\mathcal{L}_{\mathrm{gnll}}=\sum_{n,t}\frac{1}{2}\left(\log(\sigma_{n,t}^2)+\frac{|
\alpha_{\text{pred},n,t}-\alpha_{\text{true},n,t}|^2}{\sigma_{n,t}^2}\right), 
\end{equation}
where $\alpha_{\text{pred},n,t}$ and $\sigma_{n,t}^2$ are the mean and 
variance outputs of the neural network for the $n$th input trajectory at time 
$t$ \cite{nix1994estimating}. 

We train the network for 90 epochs, where each epoch constitutes a full parse 
through the training data set. For each epoch the training data set is randomly 
divided into batches of size 480, updating the network weights after each 
batch is passed through the network. Weight updates are typically achieved 
using some variant of a stochastic-gradient-descent 
algorithm~\cite{bottou2010large}. We here use 
\textit{Adam}-descent~\cite{kingma2014adam} with a cyclic learning rate 
between $1\times10^{-3}$ to $2\times10^{-5}$. During the last 10 training 
epochs, we build a so-called \textit{SWAG}-estimate~\cite{maddox2019simple} of 
the neural-network weights. These assign a Gaussian probability distribution 
to the neural-network weights by tracking their change over the training 
process. 
\textcolor{blue}{The algorithm was implemented in terms of \textit{pytorch}
\cite{paszke2019pytorch} using the \textit{NVIDIA T4} GPU with \textit{CUDA
12.0}. A full training run took roughly 50-60 hours. The run time depends
on the load of the cluster, which consisted of four such GPUs and two
\textit{AMD EPYC 7453} CPUs, allowing for up to four simultaneous runs.}

This training process was repeated multiple times resulting in a total of 6 
such \textit{SWAG}-estimates, which constitute a \textit{Multi-SWAG} estimate, 
as proposed in Ref.~\cite{wilson2020bayesian}. The final predictions on a test 
data set are then obtained by sampling 10 networks from each \textit{SWAG}-
estimate for a total of 60 Monte-Carlo-samples, which are then combined into a 
single prediction. Note that the variance between Monte-Carlo-samples is used 
as an additional, so-called \textit{epistemic}, contribution to the 
uncertainty estimate~\cite{gal2016dropout,kendall2017uncertainties}. For details on the \textit{Multi-
SWAG} procedure see Refs.~\cite{maddox2019simple,wilson2020bayesian} as well 
as Refs.~\cite{seckler2022bayesian,seckler2023machine,seckler2024multifractal} 
for example applications to anomalous diffusion. A flowchart of the described 
procedure is depicted in Fig.~\ref{fig_flowchart}.

\subsection{Evaluation metrics} 
\label{subsec_avametr}

To evaluate the performance of the network on the prediction of the full time 
series of $\alpha$, we define the \textit{mean absolute error}
\begin{equation}
    \texttt{MAE}_\alpha = \frac{1}{N} \sum_{n=1}^N \frac{1}{T} \sum_{t=0}^{T-1} |\alpha_{\text{pred},n,t}-\alpha_{\text{true},n,t}|,
\end{equation}
as well as the \textit{root mean squared error}
\begin{equation}
    \texttt{RMSE}_\alpha = \sqrt{\frac{1}{N} \sum_{n=1}^N \frac{1}{T} \sum_{t=0}^{T-1} (\alpha_{\text{pred},n,t}-\alpha_{\text{true},n,t})^2}.
\end{equation}
To visualise the performance of the error prediction we use reliability 
diagrams \cite{guo2017calibration,degroot1983comparison}, which illustrate the 
observed errors as a function of the predicted uncertainties. 
\textcolor{blue}{To achieve this performance quantification, we sort the outputs
of the neural network into sets based on a binning of the predicted standard
deviation into intervals $I_m=\bigl((m-1)\Delta_\sigma,m\Delta_\sigma\bigr]$.
We define $B_m$ as the set of output indices with a predicted standard deviation
within the interval $I_m$,
\begin{equation}
B_m = \{(i,\tau)|\sigma_{i,\tau}\in I_m=\bigl((m-1)\Delta_\sigma,m\Delta_\sigma\bigr]\},
\end{equation}
where $\sigma_{i,\tau}$ is the output standard deviation for trajectory $i$ at
time index $\tau$. For each interval $I_m$ we calculate the (observed)
\textit{root mean squared error}
\begin{equation}
\texttt{RMSE}_\alpha(B_m) = \sqrt{\frac{1}{|B_m|}\sum_{(i,\tau)\in B_m}(\alpha_{\text{pred},i,\tau}-\alpha_{\text{true},i,\tau})^2}
\label{eq_rmsebin}
\end{equation}
and compare it to the (predicted) \textit{root mean variance}
\begin{equation}
\texttt{RMV}_\alpha(B_m) = \sqrt{\frac{1}{|B_m|}\sum_{(i,\tau)\in B_m}(\sigma_{i,\tau})^2},
\label{eq_rmvbin}
\end{equation}
where $|B_m|$ is the cardinality of $B_m$. Plotting $\texttt{RMSE}$ in dependence
of $\texttt{RMV}$ for all $m$ constitutes the reliability diagram.}
Coinciding \texttt{RMSE} and \texttt{RMV} in all bins represent a perfectly
calibrated model. Deviations from the ideal error prediction can be summarised
by the \textit{expected normalised calibration error} (\texttt{ENCE})
\cite{levi2022evaluating}
\begin{equation}
\texttt{ENCE}=\sum_{B_m}\frac{|B_m|}{N
T}\frac{|\texttt{RMV}(B_m)-\texttt{RMSE}(B_m)|}{\texttt{RMV}(B_m)}.
\label{eq_ence}
\end{equation}

For change-point detection we will be given a set of predicted and true 
change-points for each trajectory. We define the gated absolute distance 
between the $i$th true change-point at time $t_{\text{cp},\text{true},i}$ and 
the $j$th predicted change-point at $t_{\text{cp},\text{pred},j}$ 
as~\cite{chenouard2014objective}
\begin{equation}
    d_{i,j} = \min(|t_{\text{cp},\text{true},i}-t_{\text{cp},\text{pred},j}|,d_{\max}),
    \label{eq_gatedabsdist}
\end{equation}
where $d_{\max}$ is the maximum penalty for change-points too far apart. The 
choice of this threshold will significantly impact the performance metric. 
Here we choose $d_{\max}=10$, aligning with the choice made in other works on 
this subject \cite{munoz2021objective,munoz2023quantitative,requena2023inferring,qu2024semantic} 
and thereby allowing for better comparability. To find the optimal pairing of 
true and predicted change-points, we need to minimise the sum of distances
\begin{equation}
    d_{\text{tot}} = \sum_{\text{pairs } i,j} d_{i,j}.
\end{equation}
Such an assignment problem can easily be solved via, for example, the 
\textit{Hungarian Algorithm} \cite{kuhn1955hungarian}. Utilising this optimal 
assignment we count the number of true positives ($\texttt{TP}$) as those 
pairs with a distance less than $d_{\max}$. Predicted change-points without 
any associated true change-point are counted as false positives 
($\texttt{FP}$), while true change-points without any paired predicted 
change-point are false negatives ($\texttt{FN}$). Paired change-points with 
distance larger than $d_{\max}$ contribute to the count of both $\texttt{FP}$ 
predicted and $\texttt{FN}$ true change-points. We calculate the total number 
of $\texttt{TP},\texttt{FP}$ and $\texttt{FN}$ over all $N$ trajectories in 
the test data set. From these we can calculate the \textit{Jaccard Similarity 
Coefficient}~\cite{jaccard1901etude}
\begin{equation}
    J = \frac{\texttt{TP}}{\texttt{TP}+\texttt{FP}+\texttt{FN}},
    \label{eq_jacc}
\end{equation}
which corresponds to the intersection of the true and predicted set of 
change-points divided by their union.
Additionally one may compute precision and recall as
\begin{eqnarray}
    \texttt{PREC} &=& \frac{\texttt{TP}}{\texttt{TP}+\texttt{FP}} \label{eq_prec}\\
    \texttt{REC} &=& \frac{\texttt{TP}}{\texttt{TP}+\texttt{FN}}. \label{eq_rec}
\end{eqnarray}
Here the first indicates how many detected change-points were correct, while the 
latter measures the share of true change-points that were detected. The 
harmonic mean of precision and recall is called the $F_1$ score,
\begin{equation}
    F_1 = \frac{2}{\texttt{PREC}^{-1}+\texttt{REC}^{-1}}.
\end{equation}
Finally, for those predictions that were detected as $\texttt{TP}$, we also 
calculate the root mean squared error
\begin{equation}
    \texttt{RMSE}_\text{CP} = \sqrt{\frac{1}{\texttt{TP}}\sum_{\substack{\text{pairs } i,j \\ d_{i,j}<d_{\max}}}(t_{\text{cp},\text{true},i}-t_{\text{cp},\text{pred},j})^2}.
\end{equation}

\section{Results}
\label{sec_results}

After training the \emph{Multi-SWAG} model, we here report the obtained 
results. First we examine the performance of the network on the sequence 
prediction of the anomalous diffusion exponent and its uncertainty as a whole 
in subsection \ref{subsec_wholeseq}. This is followed by the change-points 
extraction in subsection \ref{subsec_cpextract}, where we first ana\-lyse the 
performance when extracting change-points from only the series of anomalous 
diffusion exponent values \emph{or} the predicted variances, and later we 
explore the possibility to improve the results by a combined change-point 
determination from both. This will include a comparison of different variants 
of extraction methods. Finally, we use the sequence-predictions, now segmented 
via the change-points, to determine single-valued predictions for the two 
motion modes and report on the obtained results in 
subsection~\ref{subsec_twomodepred}.

Note that we use two test data sets for this purpose, both containing 
$N=5\times10^5$ trajectories. The first is used to determine and fine tune the 
best method to extract change-points, which is then applied to a second test 
data set in order to avoid any possible bias.

\subsection{Sequence performance}
\label{subsec_wholeseq}

\begin{figure*}
    \centering
    \includegraphics[width=\linewidth]{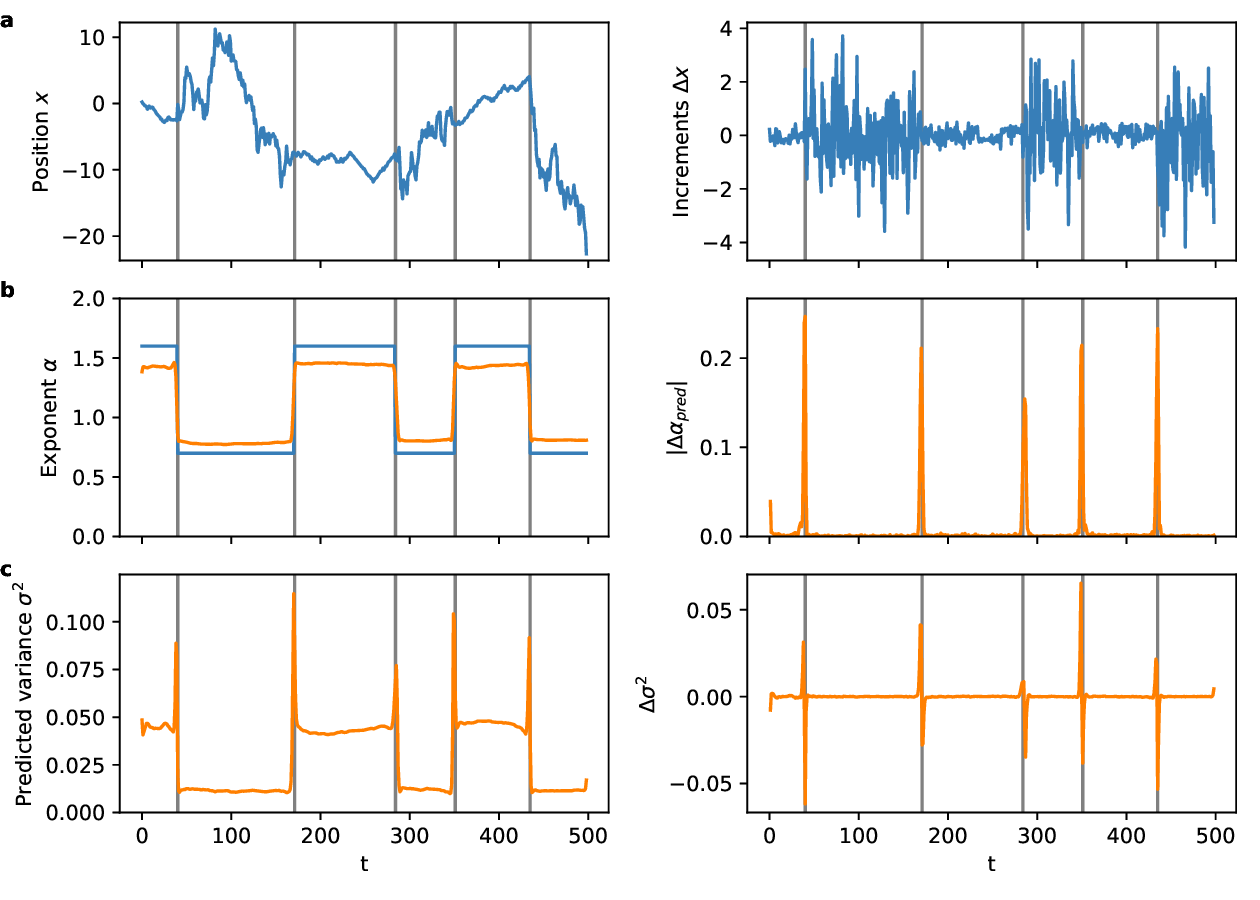}
    \caption{Sample trajectory (a) and prediction of the anomalous diffusion 
exponent (b) and uncertainty (c) given by the neural network. 
In this sample 
the trajectory changes between an LW with $\alpha=1.6$ and an
SBM with $\alpha=0.7$. The changes were generated with mean dwell 
time $\beta\approx184$.
The panels on 
the right show the changes for each of the time series a-c. In (b) the blue 
line corresponds to the true and the orange line to the predicted anomalous 
diffusion exponent. The change-points, indicated by gray lines, appear to 
coincide with maximum changes in the predicted exponent $\alpha$ or 
uncertainty $\sigma^2$. Note that the change in uncertainty in (c) features a 
double-peak-like behaviour; this is caused by a lack of confidence in the 
position of the change-point leading to a high predicted uncertainty. As such 
the presence and severity of the double peak mainly depends on the 
corresponding change in $\alpha$ compared to the base uncertainty levels 
outside of the change-point region.}
    \label{fig_examplepreds}
\end{figure*}

\begin{figure}
    \centering
    \includegraphics[width=0.9\linewidth]{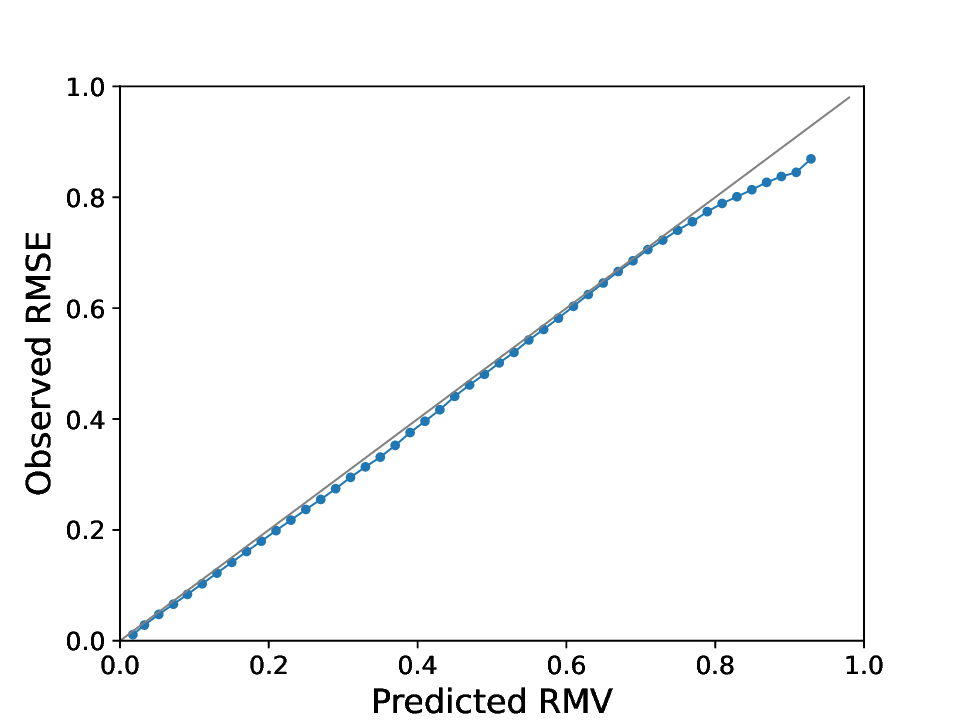}
    \caption{Reliability diagram for predictions of the whole sequence. The 
diagram shows the observed root mean squared error (\texttt{RMSE}) for given predicted 
root mean variance (\texttt{RMV}). Ideally predicted and observed errors would coincide 
as indicated by the grey line. As we see here the network slightly 
overestimates the error, this is especially prevalent for the upper end of the
predicted \texttt{RMV} and likely caused by the high errors typically assigned to the 
change-point regions.}
    \label{fig_reliability}
\end{figure}

We start by examining the network performance for the sequence prediction as a 
whole. Fig. \ref{fig_examplepreds} shows an example of an anomalous-diffusion 
trajectory and the predictions of anomalous diffusion exponent $\alpha$ and 
uncertainty $\sigma^2$ at each point in time. The panels on the right show the 
changes in each of these quantities. The figure shows that the anomalous 
diffusion exponent switches between two predictions with sharp changes at the 
change-points. Similarly the predicted variance appears to switch between two 
predictions, we here however see a peak-like behaviour at the change-points, 
which is mirrored in the change in the uncertainty (on the right of panel c) as a 
double-peak. This is caused by the uncertainty in the change-point location, 
which requires the predicted variance to account for the possibility of still 
being in either of the two modes. As such the presence and severity of this 
peak will depend on the difference in $\alpha$ between the two motion modes in 
comparison to the base uncertainty levels in each mode. Nevertheless, the 
series of uncertainties may help detect change-points, where the predicted 
exponent $\alpha$ changes little while the variance $\sigma^2$ does.

In general, when evaluating the sequence performance on the second test 
data set, we were able to achieve a mean absolute error of $\texttt{MAE}=0.178$ 
and a root mean squared error of $\texttt{RMSE}=0.259$. The magnitude of this 
error fits well within a previous study, in which the trajectories consisted 
of only one motion mode, where we were able to achieve a mean absolute error 
between $0.21$ to $0.12$ for trajectories consisting of $100$ to $500$ data 
points~\cite{seckler2022bayesian}. The large difference between $\texttt{MAE}$ 
and $\texttt{RMSE}$ is due to the increased importance placed on larger 
deviations in the $\texttt{RMSE}$. The Gaussian negative log likelihood loss, 
on which the model was trained, is $\mathcal{L}_{\mathrm{gnll}}=-1.13$. As 
this number is hard to interpret, we rely on a more explicit measure, gained 
by examining the predicted root mean variance. Fig. \ref{fig_reliability} 
shows the observed $\texttt{RMSE}$ as a function of the predicted 
$\texttt{RMV}$ in a reliability diagram. This is achieved by introducing a 
binning of the $\texttt{RMV}$ as described in Eqs. (\ref{eq_rmvbin}) and 
(\ref{eq_rmsebin}). In the figure, we observe a slightly overestimated error 
for high predicted variances. This is likely caused by the high error 
predictions associated with the change-point regions. The mean deviation 
between true and predicted errors, as defined by the expected normalised 
calibration error in Eq.~(\ref{eq_ence}), was determined as 
$\texttt{ENCE}=5.3\%$. This slight overestimation of the uncertainty is also 
mirrored when comparing the mean predicted variance on all data samples 
without binning, which yields $\texttt{RMV}=0.269$, to the observed 
$\texttt{RMSE}=0.259$.

\subsection{Change-point extraction}
\label{subsec_cpextract}

\subsubsection{Extracted from \texorpdfstring{$\alpha(t)$}{Exponent} and \texorpdfstring{$\sigma^2(t)$}{variance} independently}

\begin{figure*}
\centering
\includegraphics[width=0.9\linewidth]{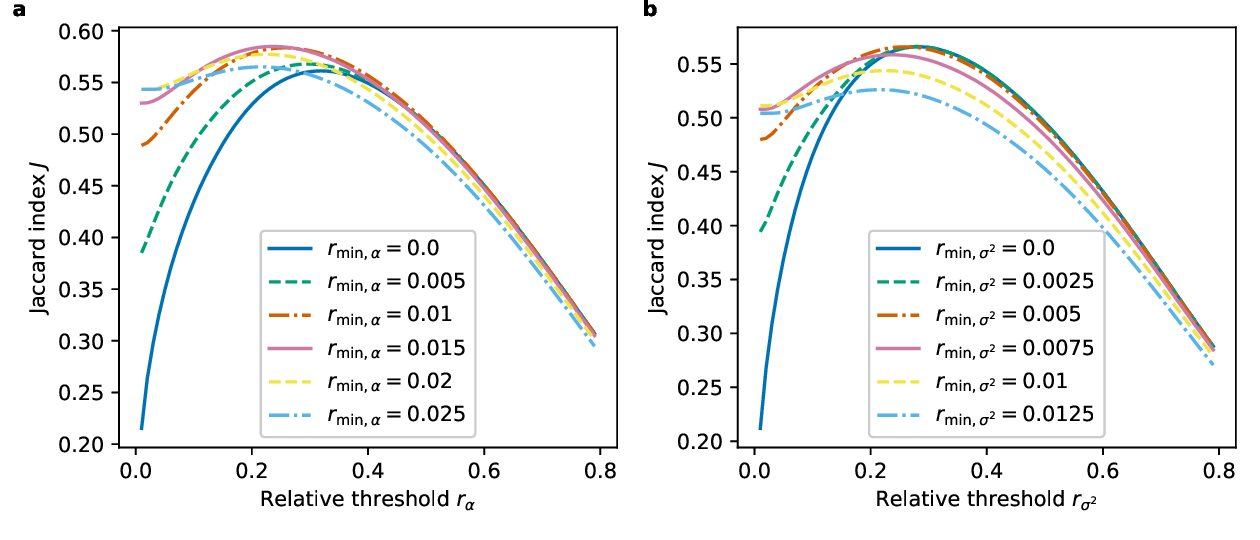}
\caption{Individual change-point performance as characterised by the 
Jaccard Index in dependence of relative and minimum thresholds. Change-points 
are extracted from only the series of anomalous diffusion exponents $\alpha$ 
(a) or the series of predicted uncertainties $\sigma^2$ (b). Change-points are 
determined via peak-detection in the increment series, the optimal thresholds 
for this detection may be inferred here. As is evident from the figure, the 
best performance is achieved when using both a threshold relative to the 
maximum observed change as well as a minimum cutoff threshold 
($r_{\alpha/\sigma^2}$ and $r_{\min,\alpha/\sigma^2}$). For $\alpha$ we find 
the best score of $J=0.585$ at a minimum threshold of $r_{\min,\alpha}=0.015$ 
and relative threshold $r_{\alpha}=0.24$. For the uncertainty the change-point 
detection appears to be more difficult, here we find a best score of $J=0.566$ at 
minimum threshold $r_{\min,\sigma^2}=0.0025$ and relative threshold 
$r_{\sigma^2}=0.28$.}
    \label{fig_thresholds_ontheirown}
\end{figure*}

\begin{algorithm}
    \caption{Change-point extraction from $\alpha(t)$}\label{alg_cpextract}
    \hspace*{\algorithmicindent} \textbf{Input:} series of predicted anomalous diffusion exponents $\alpha_0,...\alpha_{T-1}$, relative threshold $r_\alpha$, minimum threshold $r_{\min,\alpha}$, minimum distance between change-point regions $l_{\min}$\\
    \hspace*{\algorithmicindent} \textbf{Output:} vector of predicted change-points $t_{\text{cp},0},...,t_{\text{cp},K-2}$
    \begin{algorithmic}[1]
        \State{Let $\Delta\alpha_0,...,\Delta\alpha_{T-2}$ be the series of changes in $\alpha$: \newline$\Delta\alpha_t=\alpha_{t+1}-\alpha_{t}$.}
        \State{Let $\Delta_\text{thres}$ be the threshold: \newline$\Delta_\text{thres}=\max(r_\alpha\times\max_t(|\Delta\alpha_t|),r_{\min,\alpha})$.}
        \State{Let $t_{\text{exc},0},...,t_{\text{exc},S-1}$ be the time-indices of $\Delta\alpha_t$ exceeding the threshold: $|\Delta\alpha_{t_{\text{exc},s}}|\geq\Delta_\text{thres}$. Additionally add $t_{\text{exc},0}=0$ and $t_{\text{exc},S-1}=T-1$ if not already included, as otherwise the first/last change-point will be missed.}
        \State{Let $s_0,...,s_{K-1}$ be the indices of $t_{\text{exc},s}$ with minimum distance to the previous: $t_{\text{exc},s_k}-t_{\text{exc},s_k-1}\geq l_{\min}$, thereby marking the beginning of change-point regions.}
        \State{Initialise the set of change-points $t_{\text{cp},0},...,t_{\text{cp},K-2}$.}
        \For{$k=0$ to $K-2$}
            \State{$t_{\text{cp},k}=\argmax_{t,t_{\text{exc},s_{k}}\leq t<t_{\text{exc},s_{k+1}-1}}(|\Delta\alpha_t|)$}
            
        \EndFor
        \State \textbf{return} $t_{\text{cp},0},...,t_{\text{cp},K-2}$
    \end{algorithmic}
\end{algorithm}

In the previous section, we determined that the change-points can be extracted 
from the sequence predictions by detecting peaks in the changes of $\alpha$ or 
$\sigma^2$. As an example, the procedure used for change-point extraction from 
$\alpha$ is shown in Alg.~\ref{alg_cpextract}. In short, we start by finding 
all changes that exceed a certain threshold, which will result in regions of 
data points surpassing this threshold. We consider regions as separate if 
there is a minimum number of non-exceeding data points between them. Here we 
found a minimum of $l_{\min}=6$ data points between change-point regions to 
yield the best results. From each exceeding region we extract a single 
change-point. For $\alpha$ this is the point of the maximum (absolute) change 
in the region. To accommodate the often-occurring double-peak behaviour in the 
uncertainty sequence $\sigma^2$, we differentiate between two cases. If there 
is a positive and negative peak exceeding the threshold in a region, we take 
the midpoint between the two as change-point, otherwise we proceed as for 
$\alpha$.

The choice of the threshold heavily depends on the desired metric. A high 
threshold will usually result in a high precision but at the cost of recall 
(see Eqs. (\ref{eq_prec}) and (\ref{eq_rec})), while a lower threshold will 
increase recall at the cost of precision. We chose a threshold such that the 
best Jaccard similarity (see Eq.~(\ref{eq_jacc})) is achieved. 
Fig.~\ref{fig_thresholds_ontheirown} shows the change in the Jaccard index 
depending on the chosen thresholds. Here we find best results when choosing a 
threshold relative to the maximum (absolute) change in the trajectory 
$r_{\alpha/\sigma}$, which is however limited by a minimum cutoff threshold 
$r_{\min,\alpha/\sigma^2}$. The optimal thresholds, with respect to the 
Jaccard index, were determined as $r_{\min,\alpha}=0.015,r_{\alpha}=24\%$ and 
$r_{\min,\sigma^2}=0.0025,r_{\sigma^2}=28\%$. Note that we also tried using 
the sequence of predicted standard deviations $\sigma$ instead of variances 
$\sigma^2$, but found this to diminish performance.

We then apply these thresholds, determined utilising the first test data set, 
to extract change-points on the second test data set. For the $\alpha$-sequence 
we thus achieved $J = 58.4\%$, $\texttt{RMSE}_\text{CP} = 2.57$, 
$\texttt{PREC} = 81.1\%$, $\texttt{REC} = 67.7\%$, while for the uncertainty 
sequence we obtained $J = 56.6\%$, $\texttt{RMSE}_\text{CP} = 2.63$, 
$\texttt{PREC} = 75.2\%$ and $\texttt{REC} = 69.6\%$. While extracting 
change-points from the sequence of means $\alpha$ does achieve higher 
similarity than from the variances $\sigma^2$, we do see a higher recall when 
using the variance sequence. This indicates that while more of the 
change-points are found, this does however appear to come with a significant 
cost of predicted non-existent change-points, reducing precision and 
ultimately resulting in a lower Jaccard index $J$.

\subsubsection{Combined performance}

\begin{figure}
    \centering
    \includegraphics[width=\linewidth]{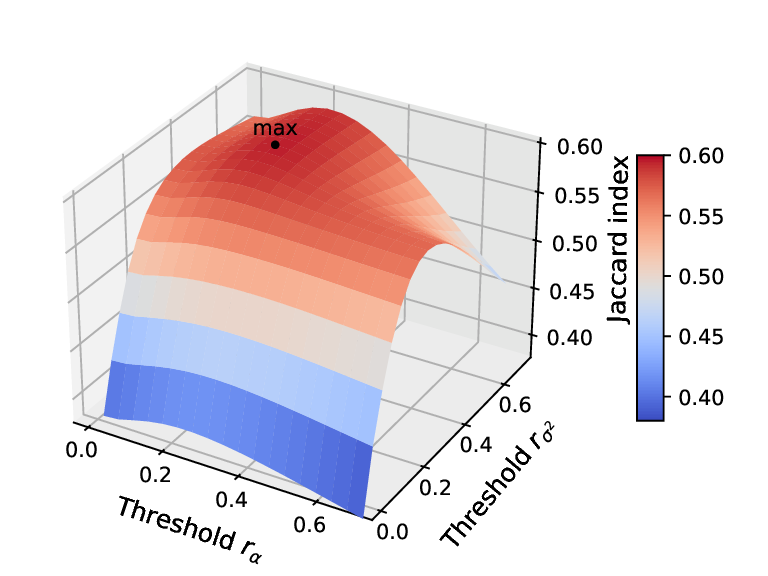}
    \caption{Combined change-point performance as characterised by the Jaccard 
index for different relative thresholds. Using the test data set we extracted 
change-points for different values of thresholds for both the series of 
anomalous diffusion exponents $\alpha$ and the series of predicted errors 
$\sigma^2$. The sets of change-points were subsequently combined into a single 
set, which was used for the calculation of the Jaccard index. Note that we 
here only show a projection from the true parameter space with set minimum 
thresholds of $r_{\min,\alpha}=0.02$ and $r_{\min,\sigma^2}=0.0015$. The best 
performance of $J=0.595$, indicated by the black dot in the figure, was found 
for relative thresholds at $r_\alpha,r_{\sigma^2} = 0.21,0.45$ 
(and $r_{\min,\alpha},r_{\min,\sigma^2}=0.02,0.0015$).}
    \label{fig_thresholds}
\end{figure}

\begin{figure*}
\centering
\includegraphics[width=\linewidth]{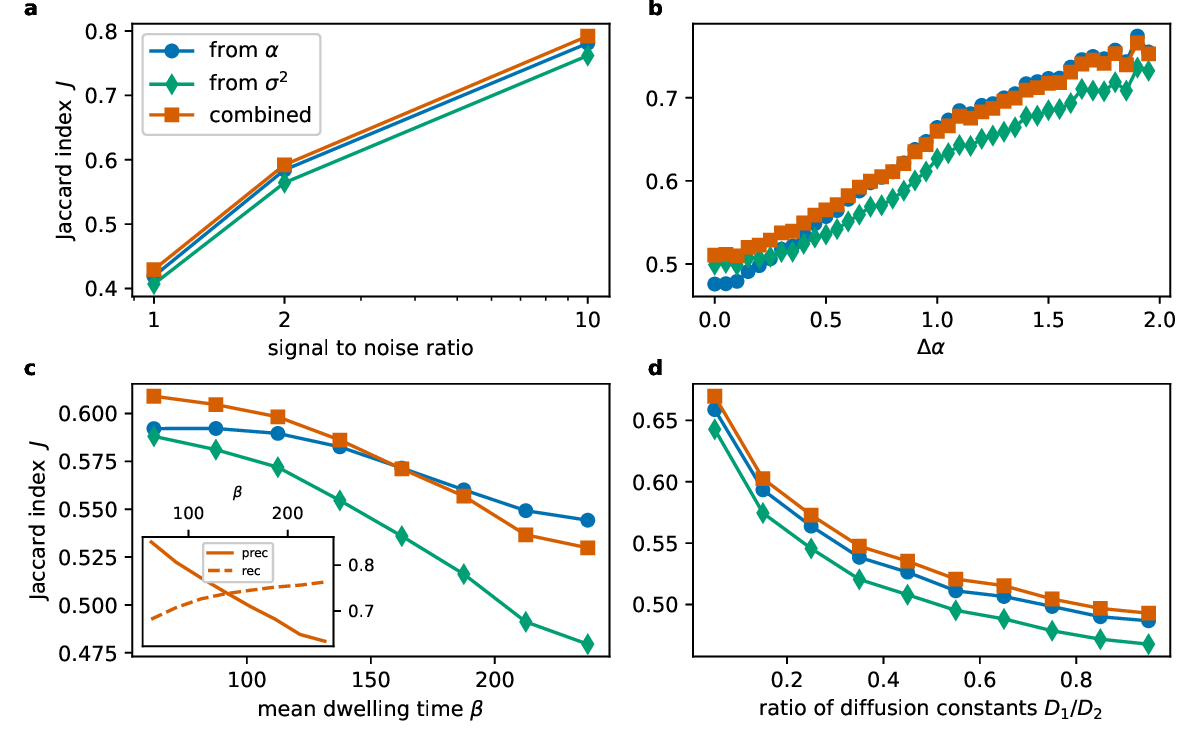}
\caption{\textcolor{blue}{Performance as characterised by the Jaccard
index in dependence of the data set generating parameters, consisting of
(a) the signal to noise strength ratios, (b) the absolute difference in
anomalous diffusion exponent $\Delta\alpha$ between the two diffusive modes,
(c) the parameter $\beta$ (equal to mean and inverse of rate) used for the
generation of dwell times (see Eq.~\ref{eq_dwell}) and (d) the ratio of
diffusion constants between the two modes. The figure shows the performance
for extracted change-points from the predicted series of anomalous diffusion
exponents (blue circles), variances (green diamonds) and the combined approach
(red squares). The inset in panel (c) shows precision (solid line) and recall
(dashed line) for the combined approach. The change-points extracted
from the $\alpha$-series consistently outperform the ones extracted from
$\sigma^2$-series, with the exception of small $\Delta\alpha$ in (b). In
most cases the combined approach improves on both other variants, with the
exception of high dwell times between change-points in (c).}}
\label{fig_parameterscore}
\end{figure*}

In this section we determine whether a combined change-point prediction from 
both series can improve the performance. To achieve this we tested a multitude 
of methods. We found that approaches to extract change-points from a combined 
series, for instance via defining exceeding data points as those for which the 
change in $\alpha$ and/or the change in $\sigma^2$ exceed a threshold, were 
inferior to an approach for which one first extracts change-points from the 
two series individually and then combines the two change-point sets. The 
combination of the two change-point sets is achieved using the Hungarian 
algorithm with gated absolute distance as in Eq.~(\ref{eq_gatedabsdist}), 
equivalent to the one used for the calculation for the Jaccard index. Paired 
change-points with distance less than the maximum are combined into their 
mean, while for others both are added into the combined set of predicted 
change-points.

Note that we cannot use the same thresholds as in the previous section, as 
this results in a Jaccard index between those obtained for the two individual 
series. Instead we again need to determine optimal thresholds, this time by 
varying all four ($r_\alpha,r_{\sigma^2},r_{\min,\alpha},r_{\min,\sigma^2}$) 
at the same time. We show a projection of this parameter space in 
Fig.~\ref{fig_thresholds}. The maximum Jaccard similarity is achieved when 
using thresholds similar to the previous ones for the $\alpha$-series 
($r_\alpha=21\%,r_{\min,\alpha}=0.02$), while using a high relative threshold 
for the series of uncertainties $\sigma^2$ 
($r_{\sigma^2}=45\%,r_{\min,\sigma^2}=0.0015$), indicating that only the more 
certain change-points of the latter are used to supplement the---mostly 
similar to the independently extracted---set of change-points from the former.

Applying the same thresholds to the second test data set, we were able to 
improve the Jaccard index to $J = 59.5\%$, $1.1$ and $2.9$ percentage points 
higher than the ones obtained for the individual series of anomalous diffusion 
exponent and uncertainty respectively. Additionally we find 
$\texttt{RMSE}_\text{CP}=2.56$, $\texttt{PREC} = 78.8\%$ and 
$\texttt{REC} = 70.8\%$.
Similar to the difference between change-points from 
only the $\alpha$- and $\sigma^2$-series, we observe an increase in recall at 
the cost of a decrease in precision---here however the decrease in precision is 
significantly lower and the increase in recall higher, ultimately leading to 
an increased Jaccard similarity. A similar effect may also be found in the 
harmonic mean of precision and recall
($F_1$ score), which increases to $F_1 = 74.5\%$ for the combined prediction 
compared to $F_1 = 73.8\%$ for change-points from the $\alpha$-series 
($F_1=72.3\%$ for $\sigma^2$-series).
While the increase in $J$ or $F_1$ is small, we see a 
substantial increase in recall by $3.1$ percentage points as compared to the 
extraction from only $\alpha$.
This can be significant, as often ensuring that change-points are found 
(recall) is more import than not detecting false change-points (precision).

\textcolor{blue}{Figure \ref{fig_parameterscore} shows the obtained
Jaccard score for all three extraction variants ($\alpha$,$\sigma^2$ and
combined) in dependence of the parameters, that were randomly chosen for the data
set generation. When varying the noise strength in panel (a) the score of
the combined approach shows a strong drop to $J=59\%,42\%$ for high noise
strengths of $\text{snr}=2,1$, with a significantly higher score of $J=79\%$
for the (typically more realistic) $\text{snr}=10$. The drop is similar,
regardless of which series was used, though the combined approach outperforms
the change-points extracted from $\alpha$ and $\sigma^2$ individually.}

\textcolor{blue}{In panel (b) we see the only point where the series of
predicted variances can outperform the change-points extracted from the series
of predicted means of $\alpha$, which occurs when the change in $\alpha$ between
the two motion modes is small. This is the area where we also expected the
inclusion of uncertainties to yield improvements. In the combined approach
this improvement is even stronger, increasing $J$ by $3.5$ percentage points
when $\Delta\alpha=0$ and also persisting for longer, up to $\Delta\alpha=0.5$,
closely following the score of the $\alpha$-series afterwards.}

\textcolor{blue}{Interestingly in panel (c), wee see that the score generally
drops with lower change-rates (higher mean dwell time $\beta$). This
is likely caused by the high cost in accuracy that occurs when a single
wrong change-point is predicted for low change-rates, which will have a
significantly less impact for high rates. This can be further understood by
examining the inset of panel (c), which shows a decreasing precision (ratio
of retrieved change-points that were correct) and an increasing recall (ratio
of true change-points that were retrieved) for the combined extraction. The
increase in $J$ for low change-rates (high $\beta$) is especially severe for
the series of uncertainties, which also appears to drag the performance of
the combined approach below the the one using only the $\alpha$-series for
high $\beta$. As this is the only case in which the combined approach shows
a significantly reduced performance, it might indicate a possible improvement
avenue for future research.}

\textcolor{blue}{The final panel (d) reveals a to-be-expected decreased
performance when the two diffusion coefficients of the diffusive modes are
closer to each other, dropping from $J=67\%$ to $J=49\%$ in the combined case. Here
again we see the typical pattern of the combined approach outperforming the
change-points extracted from the $\alpha$-series, which again outperforms
the $\sigma^2$-series.}

As a point of reference, during the second \emph{AnDi-Challenge} the top 
participants reached similarity scores from $J=65\%$ up to $J=70\%$ 
(and errors from $\texttt{RMSE}=1.63$ up to $\texttt{RMSE}=1.7$)
when identifying change-points in a more phenomenological trajectory 
setup~\cite{munoz2023quantitative}. Requena et al. were able to obtain a 
similarity of $J=51.5\%$ when detecting change-points using the anomalous 
diffusion exponent in FBM trajectories of length 
200~\cite{requena2023inferring}. In a similar setup, Qu et al. achieved an 
$F_1$ score of $F_1=88\%$ for FBM trajectories of length 500, here however all 
diffusive modes featured the same diffusion coefficient $K_\alpha=1$ and noise 
standard deviation $\sigma_n=0.125$ and were limited to a maximum of five 
change-points per trajectory~\cite{qu2024semantic}. While none of these can be 
compared directly to the results obtained in this work, they may be used as an 
indication of the magnitude to be expected for $J$ or $F_1$.

\subsection{Parameter prediction for the two modes}
\label{subsec_twomodepred}

We conclude by extracting the parameter values of the two motion modes and 
evaluating the results. Note that simply training a neural network to directly 
output single-valued predictions for the two modes is likely to yield better 
results, such a method however would not allow for change-point extraction and 
would not be easily transferable to an arbitrary number of motion modes. 
Instead we will here extract single-valued predictions 
($\alpha_1,\sigma_1^2,\alpha_2,\sigma_2^2$) for the two modes from the 
predicted time series of $\alpha(t)$ and $\sigma^2(t)$.

Each trajectory was divided into sequences based on the predicted 
change-points, and the means of $\alpha$ and $\sigma^2$ for each sequence were 
calculated. Subsequently these values were sorted into two sets based on 
proximity, and the means of each were used as final predictions of 
$\alpha_{1/2}$ and $\sigma^2_{1/2}$ for the first and second model. This 
results in an $\texttt{MAE}=0.190$ and $\texttt{RMSE}_\text{CP}=0.268$, 
slightly worse than the values obtained for the whole sequence in section 
\ref{subsec_wholeseq}. This is not surprising, as the calculations of the 
error here place equal importance on both motion modes, whereas for the whole 
sequence the mode maintained for a longer time (and thus easier to predict) 
has a larger impact on the error. Likely due to the additional uncertainty 
connected to the position of the change, the predicted error extracted in this 
manner slightly overestimates the true observed error, resulting in a 
predicted $\texttt{RMV}=0.298$ and mirrored in the calibration error of 
$\texttt{ENCE}=7.4\%$.

\section{Discussion}
\label{sec_discuss}

Recent years have seen an upsurge in machine learning solutions for the 
analysis of anomalous 
diffusion~\cite{munoz2021objective,seckler2023machine,gentili2021characterization,granik2019single,kowalek2019classification,firbas2023characterization,al2022classification,bo2019measurement,gajowczyk2021detection,argun2021classification,garibo2021efficient,li2021wavenet,verdier2021learning,kowalek2022boosting,kowalek2019classification,loch2020impact,gajowczyk2021detection,munoz2020single,seckler2024multifractal,seckler2022bayesian,munoz2020anomalous}. 
The potential of these methods was displayed during the \emph{Anomalous 
Diffusion (AnDi) Challenge}~\cite{munoz2020anomalous,munoz2021objective}, 
which already featured a segmentation task. This task however remained with a 
simple problem of a single change-point and did not see a very high 
participation. Since then applications of machine learning to segmentation 
tasks in diffusion trajectories have developed further.

While in some cases change-points may be determined without resorting to 
machine learning~\cite{wagner2017classification,persson2013extracting,stanislavsky2024fractional,janczura2021identifying}, 
this can prove difficult for more subtle changes. In such cases, recent works 
indicate that a point-wise prediction of anomalous diffusion parameters using 
machine learning may be a promising 
solution~\cite{requena2023inferring,qu2024semantic}. Detecting and specifying 
change-points of diffusion trajectories is also the topic of the 
just-completed 
\emph{2\textsuperscript{nd} AnDi-Challenge}~\cite{munoz2023quantitative}.

In a previous paper~\cite{seckler2022bayesian} we discussed the potential of an 
added uncertainty prediction when determining parameters of a single-particle 
anomalous-diffusion trajectory using a neural network. This is expanded upon 
in the current work, where we explored the applicability of these uncertainty 
outputs for change-point detection. For this purpose, we generated data sets of 
trajectories switching between two diffusive modes. Through \emph{Bayesian 
deep learning}~\cite{maddox2019simple,wilson2020bayesian}, we were able to 
obtain a point-wise prediction not only of the anomalous diffusion exponent 
$\alpha$ but also of the corresponding uncertainty, as measured by the 
variance $\sigma^2$. In Fig.~\ref{fig_examplepreds} we saw that this variance 
shows an interesting behaviour at the change-points, often resulting in distinct
spikes at the change-point position. We hypothesised, that the uncertainty 
series may help detect change-points where the predicted exponent changes 
little, as those would be hidden in the $\alpha$-series.

While in the present study we found that extracting the change-points 
from these uncertainties does 
indeed lead to more true positive change-point determinations, improving 
recall, it struggles in precision, adding many false positive change-points as 
well. Ultimately change-points extracted from only the uncertainty-series 
performed worse than those extracted from the series of anomalous diffusion 
exponents, when measured by the Jaccard index. However, when utilising a 
combined change-point extraction from both series we were able to improve the 
results, significantly increasing recall with a smaller loss in precision, 
which resulted in an improvement of the Jaccard index from $J=58.4\%$ to 
$J=59.5\%$. Additionally we showed that the parameters of the two motion modes 
can be recovered from the predicted sequence and change-points, albeit with a 
small loss in accuracy.

An extension of the discussed methods to trajectories switching between more 
than two modes may be of interest for the future, though the methods discussed 
here were intentionally chosen to be easily transferable to an arbitrary 
number of modes. Similarly an application to different anomalous diffusion 
models, like, for instance, the more phenomenological models used in the 
\emph{2\textsuperscript{nd} AnDi-Challenge}~\cite{munoz2023quantitative}, 
could yield notable results. The challenge considered predictions of not only 
the anomalous diffusion exponent but also the diffusion coefficient and model. 
Therefore examining the impact of uncertainty predictions for change-point 
detection in such cases, when more than one parameter is to be predicted, may 
be of interest. Similarly the use of uncertainty predictions in a 
classification task could be considered. Here however the difference to 
standard classification approaches would be small, as these already include 
class probabilities, even if those are often not very well calibrated due to 
overfitting~\cite{guo2017calibration,naeini2015obtaining,maddox2019simple}.
Another challenging point to include is dynamical error \cite{dynamic,seckler2023machine}.

The simple peak-detection method used here for change-point detection could 
also be reconsidered. A more involved method like, for example, Bayesian 
inference may be able to achieve better results. While Bayesian inference has 
already been applied directly to diffusion 
trajectories~\cite{thapa2018bayesian,thapa2022bayesian,park2021bayesian,munoz2021objective} 
and an application to change-point detection is being worked 
on~\cite{thapa2024baycp}, it is known to struggle when competing with machine 
learning approaches due to its high computational cost. An application to the 
point-wise predictions obtained from a machine learning algorithm however 
should incur a significantly smaller cost.

Finally one may also wish to test the limitations of the algorithms discussed 
here. Previous works have shown that a blind application of machine learning 
will often lead to erroneous predictions~\cite{seckler2023machine}. Observing 
if change-points could still be extracted, even if the predicted values in the 
sequence are sub-optimal, will be interesting.

\textcolor{blue}{Such a consideration would also shed light on the
to-be-expected behaviour for experimental data. The prediction of
change-points can there be utilised to segment trajectories showing changing
behaviour, such as observed, for example, in the movement patterns of
animals during foraging, a cell during feeding, as an effective behaviour
in an heterogeneous environment or due to transient interactions with other
objects~\cite{vilk2022unravelling,yin2018detection,saha2015diffusion,bag2015plasma,
prrmeyer,diego,sabri,
low2011erbb1,moldenhawer2022spontaneous, monasson2014crosstalk}. A properly
adjusted algorithm may even be applied to find change-points in financial
time series, which often show changing behaviour due to corresponding
real world decisions~\cite{malkiel1999random,plerou2000economic}. Once
segmented the predictions of the machine can be easily double-checked using
the more traditional methods, such as the mean squared displacement, which
can otherwise be hard to apply to heterogeneous trajectories.}

\begin{acknowledgments}
We thank the German Ministry for Education and Research (NSF-BMBF project 
STAXS) and the German Research Foundation (DFG, grant no. ME 1535/12-1). 
\end{acknowledgments}

\end{document}